\documentstyle[12pt]{article}
\newcommand{\beq}{\begin{equation}}
\newcommand{\eeq}{\end{equation}}
\newcommand{\bea}{\begin{eqnarray}}
\newcommand{\eea}{\end{eqnarray}}
\newcommand{\T}{\textstyle}
\newcommand{\D}{\displaystyle}
\newcommand{\tr}{{\rm tr}}
\newcommand{\re}{{\rm Re}}
\newcommand{\V}{{\cal V}}
\newcommand{\vev}[1]{\Big\langle #1 \Big\rangle}
\input epsf
\begin{document}

\hfill \vbox{\hbox{UCLA/00/TEP/06}
             \hbox{INLO-PUB 02/00}} 
\begin{center}{\Large\bf Computation of the Vortex Free Energy in    
SU(2) Gauge Theory 
}\\[2cm] 
{\bf Tam\'as G. Kov\'acs} \\
{\em  Department of Physics, Instituut-Lorentz for Theoretical Physics, 
P.O.Box 9506, 2300 RA, Leiden, The Netherlands}\\
{\sf e-mail: kovacs@lorentz.leidenuniv.nl}\\[5mm] 

and\\[5mm]

{\bf E. T. Tomboulis} \\
{\em Department of Physics, UCLA, Los Angeles, 
CA 90095-1547}\\
{\sf e-mail: tombouli@physics.ucla.edu}
\end{center}
\vspace{1cm}

\begin{center}{\Large\bf Abstract} 
\end{center}

We present the first measurement  of the 
vortex free-energy order parameter at weak coupling for SU(2) 
in simulations employing multihistogram methods. 
The result shows that the excitation probability for a  
sufficiently thick vortex in the vacuum   
tends to unity. This is rigorously known   
to provide a necessary and sufficient 
condition for maintaining confinement at weak coupling 
in $SU(N)$ gauge theories.

\vfill
\pagebreak

The vortex free energy (also known as magnetic-flux 
free energy) order parameter in gauge theories is 
defined as the ratio of the partition function in 
the presence of a topologically trapped vortex excitation 
(introduced by a singular gauge transformation) to that 
without it. Its Fourier transform w.r.t. to the center ($Z(N)$) 
of the gauge group ($SU(N)$) defines the so-called 
electric-flux free energy which is rigorously known to 
provide an upper bound on the 
Wilson loop. These flux order parameters 
can characterize all possible phases of a (pure) 
gauge theory, and furthermore do this in terms of the 
behavior of the excitation expectation for a vortex. 
They were first considered in the study of gauge theories 
in \cite{tH}, though the use of the analogous quantities 
in statistical mechanics goes back much further \cite{On}. 
The idea that vortex configurations underlie confinement at 
weak coupling has a long history, and has been the subject of 
intense recent activity. (We refer to Ref. \cite{KT} for 
a review of recent developments and references 
to early and recent work.)  

In view of the physical significance of the magnetic-flux 
free energy, it may appear surprising that it has not been 
measured in simulations over the last twenty years. 
Accurate determination of (differences of) free energies 
in gauge theories, however, is well-known to be difficult. 
In fact, it is at first not quite clear how one should go 
about computing such totally nonlocal (lattice-length) 
quantities. We present here a computation for the group 
$SU(2)$ based on multihistogram methods \cite{FS}. Such a 
method was recently used in Ref. \cite{HRR} to compute  
the free energy of a pair of $Z(N)$ monopoles, 
a quantity related to the 't Hooft loop operator. 
Our result demonstrates that the excitation expectation 
for a sufficiently extended `thick' vortex at large 
$\beta$ is essentially unity. This is the feature responsible 
for maintaining the confining phase in $SU(N)$ gauge theories 
even at weak coupling.

We work on a $d$-dimensional hypercubic lattice $\Lambda$ 
of size $L_1\times\cdots\times L_d$ with periodic boundary 
conditions in all directions. We generally denote 
bonds by $b$, plaquettes by $p$, cubes by $c$,  etc. 
The plaquette action is denoted by $A_p(U_p)$, where, as usual,  
$U_p = \prod_{b\in p}U_b$, the product of the bond 
variables around the plaquette; for the 
minimal (Wilson) action $A_p(U_p)= -\beta \re\,\tr 
U_p$. The trace "tr" is defined to include a $1/N$ 
normalization.

\begin{figure}[htb]
\begin{minipage}{155mm}
{\ }\hfill\epsfysize=4cm\epsfbox{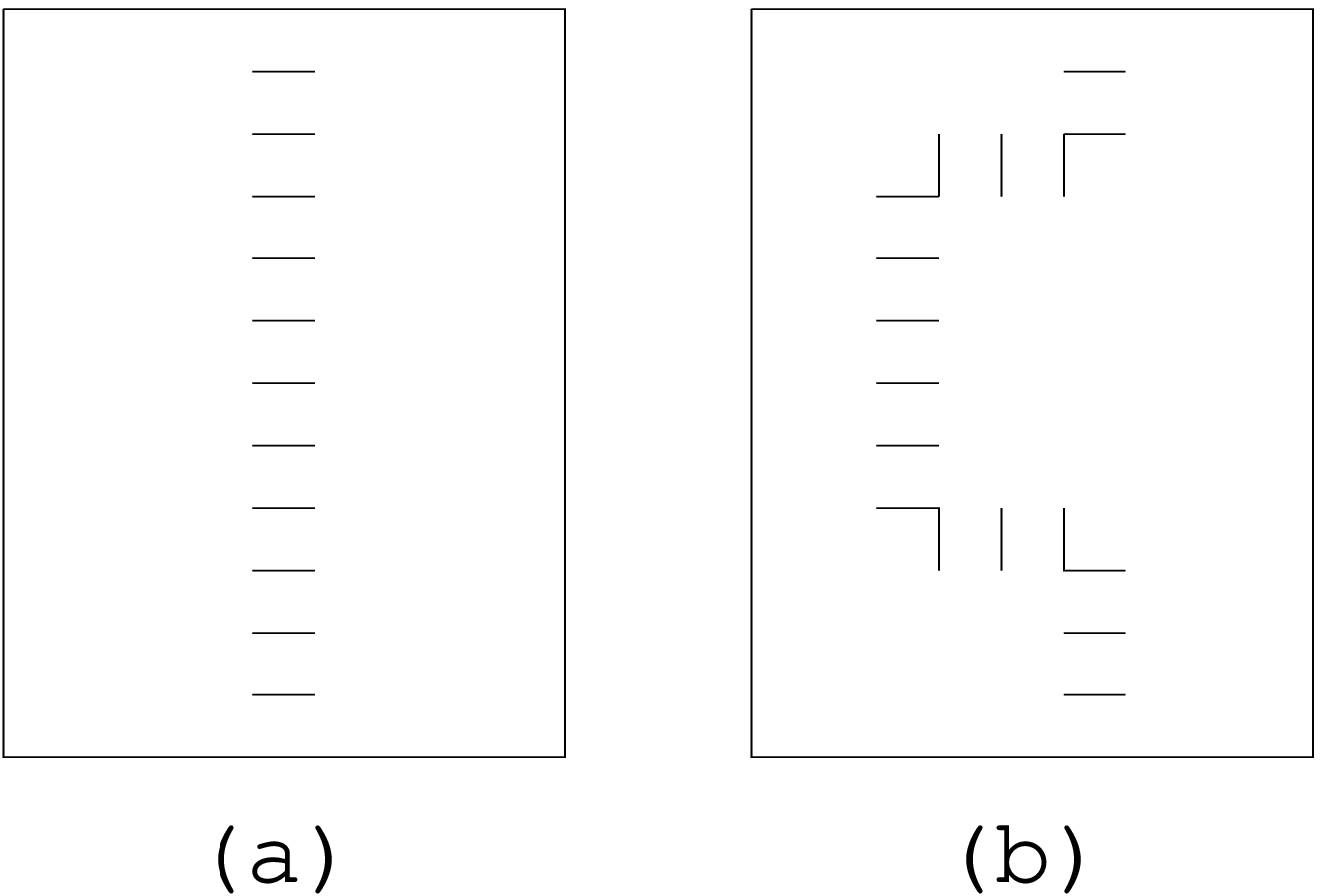}\hfill{\ }\\
\caption[twist]{\label{v1} Stack of plaquettes carrying twist 
winding around the periodic lattice. (a) and (b) are equivalent 
sets.  }
\end{minipage}
\end{figure}

A coclosed set of plaquettes (2-cells) is a closed set 
of $(d-2)$-cells on the dual lattice. Thus, in $d=3$, 
it is a closed loop of dual bonds; in $d=4$, a closed two-dimensional 
surface of dual plaquettes. For fixed $\mu$, $\nu$, let 
$\V_{\mu\nu}$ denote a coclosed set of plaquettes that 
winds through every 2-dim $[\mu\nu]$-plane of $\Lambda$, 
i.e. a topologically nontrivial plaquette set wrapped 
around the periodic lattice ($d$-torus) in the $(d-2)$ 
directions $\lambda \neq \mu,\ \nu$ perpendicular 
to $\mu,\ \nu$ This is depicted in figure \ref{v1}(a), 
where the short lines represent the plaquettes in $\V$, with  
the horizontal axis representing the $x^\mu$, $x^\nu$ directions,  
and the vertical axis the remaining $(d-2)$ perpendicular directions.

Define the partition function 
\beq
Z_\Lambda(\tau_{\mu\nu}) = \int\prod_b dU_b\;\exp\bigg(
\,- \sum_{p\not\in \V_{\mu\nu}} A_p(U_p) 
 - \sum_{p\in \V_{\mu\nu}} A_p(\tau_{\mu\nu} U_p)\,\bigg) 
\label{tPF}\,,
\eeq 
where the plaquette action $A_p(U_p)$ is replaced by the 
`twisted' action $A_p(\tau_{\mu\nu}U_p)$ for each plaquette 
of $\V_{\mu\nu}$. Here the `twist' $\tau_{\mu\nu} \in Z(N)$ 
is an element of the center. There are thus $(N-1)$ different 
nontrivial choices for $\tau_{\mu\nu}$. The trivial 
element $\tau_{\mu\nu} =1$ is the ordinary partition 
function $Z_\Lambda(1) \equiv Z_\Lambda$. 

\begin{figure}[htb]
\begin{minipage}{155mm}
{\ }\hfill\epsfysize=3.5cm\epsfbox{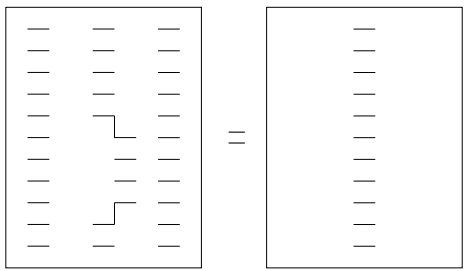}\hfill{\ }\\
\caption[tmod]{\label{v2} Equivalent sets $\V$ reflecting 
mod $N$ conservation of the twist ($N=2$). }
\end{minipage}
\end{figure}

As indicated by the notation on the l.h.s. of (\ref{tPF}),  
the exact position or shape of $\V_{\mu\nu}$ is irrelevant; 
the only dependence is on the presence of the $Z(N)$ flux 
winding through each $[\mu\nu]$-plane. 
It is indeed easily seen that $\V_{\mu\nu}$ can be moved 
around and distorted by a shift of integration variables, but 
not removed; it is rendered topologically stable by 
winding completely around the lattice (figure \ref{v1}(b)). 
By the same token introducing two twists, $\tau_{\mu\nu}$ on 
$\V_{\mu\nu}$ and $\tau_{\mu\nu}^\prime$ on $\V_{\mu\nu}^{\,\prime}$ 
in (\ref{tPF}), is equivalent to introducing one twist 
$\tau_{\mu\nu}^{\prime\prime}=\tau_{\mu\nu}
\tau_{\mu\nu}^{\prime}$ since $\V_{\mu\nu}$ and 
$\V_{\mu\nu}^{\,\prime}$ can be brought together by a shift of 
integration variables (figure \ref{v2}). This expresses the mod $N$ 
conservation of the $Z(N)$ flux introduced by the twist. Thus, 
for $N=2$, any odd number of such (nontrivial) twists is 
equivalent to one, and any even number to none.  

The magnetic-flux 
free energy order parameter is now defined as 
\bea
\exp(-F_{\rm mg}(\tau_{\mu\nu}))
& = & {Z_\Lambda(\tau_{\mu\nu})\over Z_\Lambda} \nonumber\\
&  = & \vev{\exp\bigg(\,
- \sum_{p\in \V_{\mu\nu}} \Big(\,A_p(\tau_{\mu\nu}U_p)-A_p(U_p)\,
\Big)\,\bigg)}
\;.\label{mgfe}
\eea

Generalizations of (\ref{tPF})-(\ref{mgfe}) 
may be considered by introducing sets $\V_{\kappa\lambda}$ for 
several or all of the ${1\over2}d(d-1)$ possible {\it distinct}  
choices of planes $[\kappa\lambda]$.

The twist amounts to a discontinuous (singular) $SU(N)$ gauge 
transformation on the configurations in (\ref{tPF}) 
with multivaluedness in $Z(N)$ (so  
it is single-valued in $SU(N)/Z(N)$)), i.e. the introduction of a  
$\pi_1(SU(N)/Z(N)) = Z(N)$ vortex. The set 
$\V_{\mu\nu}$ represents the topological obstruction to 
having singlevaluedness everywhere. (\ref{tPF}) is then the 
partition sum for the system with a topologically stable 
vortex completely winding around the lattice; and (\ref{mgfe}) 
is the normalized expectation for the excitation 
of such a vortex. Hence, it is also referred to as the 
vortex free energy. 

Choosing, say, $[\mu\nu]=[12]$ in (\ref{tPF}), we now drop the 
$\mu\nu$ subscript. One is interested in the behavior 
of (\ref{mgfe}) in the 
large volume limit (in the van Hove sense), i.e. as 
the size of the lattice increases in any power law fashion, 
e.g. $L_\mu=2^{la_\mu}$ for some fixed 
choice of positive exponents $a_\mu$, integer $l\to \infty$. 
Let $A=L_1L_2$ be the area of each  
$[12]$-plane, and $L=L_3\cdots L_d$ the lattice volume in the 
perpendicular directions. One is interested, in particular, 
in $L\geq A$.  
The twist introduces a cost in action localized on the 
plaquettes in $\V$. This cost, proportional to $L$, 
may be lowered if there are configurations that 
contribute with finite measure in the integral (\ref{mgfe}), 
and allow the flux introduced by the 
twist to spread in the two directions perpendicular 
to $\V$, so that the action is 
closer to its minimum;  
in other words, if there is finite probability 
for exciting a `thick' vortex.  

For sufficiently large 
lattices, there are then three possibilities ($\tau\neq 1$):  
\begin{enumerate}
\item[(a)] \ $\exp(-F_{\rm mg}(\tau)) 
                 \sim \exp(-\alpha(\beta, \tau)\,L)$
\item[(b)] \ $\exp(-F_{\rm mg}(\tau)) 
                \sim \exp(-\beta\,c(\tau)\,\D{L\over A})$ 
\item[(c)] \ $\exp(-F_{\rm mg}(\tau)) 
                  \sim \exp(-cL\,e^{-\rho(\beta,\tau)\,A})$
\end{enumerate}
In case (a) the magnetic flux stays focused in a thin vortex; 
this describes a Higgs phase. In (b) the flux can spread in 
a Coulomb-like fashion lowering the free-energy cost; this 
describes a massless Coulomb phase, where the long distance 
behavior is accurately given by weak coupling pertubative 
expansion. In (c) the gain in thickening the vortex 
is exponential;  this characterizes the confinement phase. 
It is important to note that, 
in contrast to (a)-(b), only (c) gives a value which    
survives and in fact tends (exponentially) to unity for all  
ways of taking the thermodynamic limit as described above; 
this is the signature of the confinement phase.

Since our computation below is for $N=2$,  
we now write explicit formulae only for this case.  
The Fourier transform of (\ref{mgfe}) w.r.t. $Z(N)$ is 
known as the electric-flux free energy. For $N=2$ this is 
simply:  
\beq 
\exp(-F_{\rm el}) = \sum_{\tau=1,-1}\; 
\;\tau\;\exp(-F_{\rm mg}(\tau))  
=1 - \exp(-F_{\rm mg}(-1)). 
   \label{elfe}
\eeq

Consider now a rectangular loop $C$ in a $[12]$-plane.   
Then, for any reflection positive plaquette action, 
the Wilson loop obeys the bound \cite{TY}:
\beq 
\vev{\,\tr (U[C])\,} \leq \bigg(\,\exp(-F_{\rm el}) 
\,\bigg)^{\T{A_C\over A}} \label{bWL}\;,  
\eeq 
where $A_C$ is the minimal area bounded by $C$.  
(\ref{bWL}) shows that confining behavior (c) for the vortex 
free energy implies area-law for the Wilson loop with string 
tension bounded from below by the excitation expectation 
for a vortex. So confining behavior for the vortex free energy 
is a {\it sufficient condition} for linear asymptotic quark 
confinement. 

Placing suitable constraints in the functional measure in 
(\ref{tPF}) which forbid the spreading of flux across 
$[12]$-planes, thus eliminating the occurance of thick vortices, 
results in nonconfining behavior of type (a) above \cite{Y}.  
In this case (\ref{bWL}) cannot tell us anything about the 
Wilson loop. To show loss of confining behavior for the Wilson 
loop itself in the presence of such constraints, one needs 
a {\it lower} bound on it which exhibits perimeter-law. 
This was recently proven for large $\beta$ in \cite{KT1}. 
Thus the occurance of thick vortices is also a {\it necessary 
condition} for confinement at weak coupling.

Our measurement of (\ref{mgfe}) for $SU(2)$ 
was done by an application of a multihistogram method 
\cite{FS}. From now on we restrict the form of the action
to the Wilson action
\beq
  A_p(U_p) = -\beta \tr U_p \; ,
\eeq
which was used in the measurement. The 
basic quantity in our procedure is  
the density of states $w(S)$ as a function of the total 
action $S$ along the twisted plaquettes. This is defined as 
\beq
  w(S) = \prod \int dU_b \; 
         \exp\Big( \beta \sum_{p\notin\V} \tr U_p\Big)\;
         \delta(S+\sum_{p\in\V} \tr U_p )\;\,.
\eeq
If $w(S)$ is known, the partition function can be easily
computed for any coupling $\beta_\V$ along $\V$ as
\beq
  Z(\beta_\V) = \int dS \; w(S) \; e^{-\beta_\V S} \;.
\eeq
In particular, we are interested in $Z(\beta)$, the
untwisted, and $Z(-\beta)$, the twisted partition function.
The problem is that the dominant contribution for
$Z(\beta_\V)$ comes from different regions of $S$, depending
on $\beta_\V$. Therefore one needs to know $w(S)$ to a good
accuracy in a wide range of $S$. A simulation done at a
certain value of $\beta_\V$, however will give accurate 
information on $w(S)$ only in a narrow neighbourhood of 
$\langle S \rangle_{\beta_\V}$. The main idea of the
Ferrenberg-Swendsen multihistogram method is to combine
information on $w(S)$ coming from simulations at different
$\beta_\V$'s to obtain $w(S)$ in a wide range of $S$ 
accurately. This can be done by noting that for a given
$\beta_\V$ the probability distribution of $S$, $P(S,\beta_\V)$,
goes as
\beq
  P(S,\beta_\V) \propto \frac{1}{Z(\beta_\V)} \; 
  e^{-\beta_\V S}\;,
\eeq
and that $P(S,\beta_\V)$ can be directly measured by making
a histogram of the action along $\V$. In this way, 
any simulation at a certain $\beta_\V$ gives an estimate for $w(S)$,
\beq
   w(S) = P(S,\beta_\V) \; e^{\beta_\V S} \;
          Z(\beta_\V)\;.
\eeq
These estimates coming from simulations with different
$\beta_\V$'s (say $\beta_1$, $\beta_2$,...$\beta_K$) 
can then be averaged with suitable ($S$ dependent)
weights to minimise the error in $w(S)$ over a given range
of $S$. This results in the following set of coupled equations: 
\begin{eqnarray}
  w(S) = \frac{\sum_{n=1}^K \; P(S,\beta_n)}{\sum_{n=1}^K
               \frac{\T\exp(-\beta_n S)}{\T Z(\beta_n)}} \\
  Z(\beta_n) = \int dS \; e^{-\beta_n S} \; w(S) \;,
\end{eqnarray}
which can be solved by iteration starting from 
$Z(\beta_n)=1$. To optimize the procedure, one needs
a sufficient overlap between the $P(S,\beta_n)$ 
distributions corresponding to successive $\beta_n$'s.
Since the distributions quickly become narrower
with increasing lattice size, the number of simulations,
$K$, also needs to be increased accordingly. This makes
our measurement very expensive on large lattices. 
For the largest lattices we
typically used $K=40-80$ simulations with the 
$\beta_n$'s equally spaced in the $-\beta$ $+\beta$
range.

\begin{figure}[htb]
\begin{minipage}{155mm}
{\ }\hfill\epsfysize=7cm\epsfbox{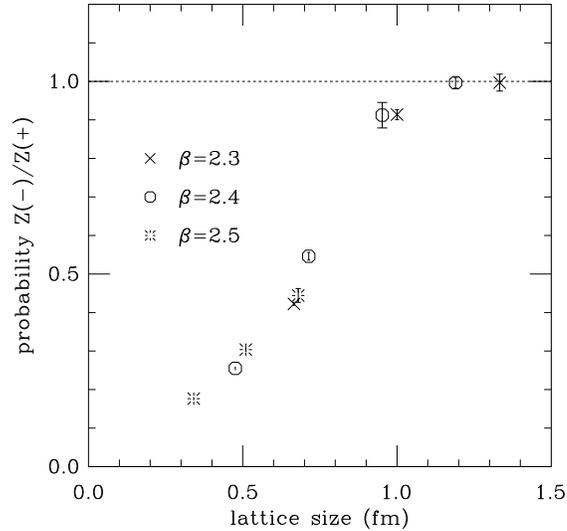}\hfill{\ }\\
\caption[tagv]{\label{vfe}$SU(2)$ vortex free energy 
(\ref{mgfe}) vs. lattice size }
\end{minipage}
\end{figure}

The result of the computation for (\ref{mgfe}) is 
shown in figure \ref{vfe}. We have performed the computation 
on lattices of equal linear size in all directions 
for three different values of $\beta$. The lattice spacings are 
$a=0.165$ fm, $a= 0.119$ fm  and $a=0.085$ fm for $\beta=2.3$, 
$\beta=2.4$, and $\beta=2.5$, resp. 

Notice that, with 
the lattice size expressed in physical units, the 
measurements for different $\beta$'s fall on the same curve, 
as they should. This indicates that the universal curve 
has been reached, and will not change at larger beta. 
Also, the onset of the sharp rise around $0.7$ fm is 
in the region of the finite temperature deconfining phase 
transition providing another indirect consistency check. 

The approach to unity for sufficiently large lattice size 
in figure \ref{vfe} is striking. In comparison, for 
Coulomb-like massless behavior, an upper bound obtained by 
action minimizing within the spin-wave approximation gives 
$\sim \exp(\,-\beta\,(\pi/2)^2\,) \approx 0.085$ at 
$\beta=2.3$. The points forming the upper part of the plot  
are well within the confinement region. 
The string tension values extracted from the vortex free energy
in the confining region are consistent with the 
values from heavy-quark potential calculations 
(see eg.\ \cite{FHK}), though still better precision in  
the measurement of the vortex free energy is required 
for precise quantitative comparisons.  
     
In conclusion, the result of our computation clearly 
demonstrates that the weighted expectation for 
the excitation of a sufficiently thick vortex in the 
vacuum tends to one. In this sense the vacuum can indeed 
be viewed as having a `condensate' of thick long vortices. 
This is sufficient for maintaining confinement at large 
$\beta$ in $SU(N)$ gauge theories. As mentioned, rigorous 
results also show it to be necessary: were the behavior for 
(\ref{mgfe}) exhibited in figure \ref{vfe} not to occur, 
confinement at large beta would be lost.    

We are very grateful to P.\ de Forcrand for correspondence.
This research was supported by FOM (T.G.K.) and by NSF, 
Grant No.\ NSF-PHY-9819686 (E.T.T.).

\end{document}